\documentclass[conference]{IEEEtran}
\usepackage{fancyhdr} 
\IEEEoverridecommandlockouts
\usepackage{cite}
\usepackage{amsmath,amssymb,amsfonts}
\usepackage{algorithm}
\usepackage{algpseudocode}
\usepackage{graphicx}
\usepackage{textcomp}
\usepackage{xcolor}
\def\BibTeX{{\rm B\kern-.05em{\sc i\kern-.025em b}\kern-.08em
    T\kern-.1667em\lower.7ex\hbox{E}\kern-.125emX}}

\usepackage{url}

\usepackage{multirow}
\usepackage{multicol}
\usepackage{hhline}

\usepackage{listings}
\usepackage{subcaption}

\usepackage{adjustbox}

\usepackage{xspace}
\usepackage{todonotes}
\usepackage{comment}

\newcommand{\wse}{WSE\xspace}

\begin{document}

\title{
Matrix-Free Finite Volume Kernels\\ on a Dataflow Architecture
}

\author{\IEEEauthorblockN{
Ryuichi Sai\IEEEauthorrefmark{1},
Fran{\c c}ois P. Hamon\IEEEauthorrefmark{2},
John Mellor-Crummey\IEEEauthorrefmark{1},
Mauricio Araya-Polo\IEEEauthorrefmark{2}
}
\IEEEauthorblockA{\IEEEauthorrefmark{1}Rice University, Houston, TX, USA}
\IEEEauthorblockA{\IEEEauthorrefmark{2}TotalEnergies EP Research \& Technology US, LLC., Houston, TX, USA}}

\maketitle

\begin{abstract}
Fast and accurate numerical simulations are crucial for designing large-scale geological carbon storage projects ensuring safe long-term CO\textsubscript{2} containment as a climate change mitigation strategy.
These simulations involve solving numerous large and complex linear systems arising from the implicit Finite Volume (FV) discretization of PDEs governing subsurface fluid flow.
Compounded with highly detailed geomodels,
solving linear systems is computationally and memory expensive, and accounts for the majority of the simulation time.
Modern memory hierarchies are insufficient to meet the latency and bandwidth needs of large-scale numerical simulations.
Therefore, exploring algorithms that can leverage alternative and balanced paradigms, such as dataflow and in-memory computing is crucial.
This work introduces a matrix-free algorithm to solve FV-based linear systems using a dataflow architecture to significantly minimize memory latency and bandwidth bottlenecks.
Our implementation achieves two orders of magnitude speedup compared to a GPGPU-based reference implementation, and up to 1.2 PFlops on a single dataflow device.
\end{abstract}

\begin{IEEEkeywords}
finite-volume,
matrix-free linear solver,
high-performance computing,
energy,
dataflow architecture,
wafer-scale engine,
distributed memory
\end{IEEEkeywords}

\section{Introduction}
Global warming due to high CO\textsubscript{2} concentrations in the atmosphere is of growing concern. As a result, there is increasing interest in mitigation strategies that reduce atmospheric CO\textsubscript{2}.
The Intergovernmental Panel for Climate Change has identified Geological Carbon Capture and Storage (CCS) as a potential mitigation strategy~\cite{srccs_report}.
To meet the United Nations targets~\cite{srccs_report} for reducing CO\textsubscript{2} production, it is necessary deploy CCS capabilities at a large scale.
Designing large-scale CCS projects requires numerical simulations to ensure safety and long-term containment of the CO\textsubscript{2}.
Fast and accurate simulation is crucial to designing safe projects within regulatory and commercial time constraints.

Numerical simulations of physical processes involved in CCS require solving complex PDEs that govern subsurface fluid flow.
Current commercial and research simulators use implicit Finite-Volume (FV) schemes with Two-Point Flux Approximation (TPFA) to discretize the governing equations.
On detailed geomodels, such simulations are computationally and memory intensive.
On traditional HPC systems, the performance of such simulations is limited by (host and device) memory hierarchy and networking performance~\cite{reed2022reinventing}.
Therefore, it is important to explore implementation of such methods on alternative architectures, e.g. dataflow chips, and corresponding algorithms to enable fast FV-based flow simulations.

Advancements in HPC system design have enabled optimizations and algorithmic changes in scientific and business fields.
Previous investigations into non-hierarchical architectures have improved computational efficiency~\cite{MauricioIBMCellBE,kahle1989connection}.
The emergence of highly parallel systems with distributed memory architecture is now considered as alternative to traditional accelerated systems.
New chips from Cerebras~\cite{cerebras_hotchips2019}, Groq~\cite{groq_HotChip}, and SambaNova~\cite{Samba21}, systems with dataflow-like architecture design and on-chip memory possess higher memory bandwidth, lower memory latency, and lower energy cost for memory access.

In this article, we explore the capabilities of a dataflow architecture system by implementing a matrix-free solution method for single-phase problems that rely on localized communications and single-level memory.
The proposed advancements have the potential to benefit reference CCS simulators such as GEOS \cite{GEOSXWebsite,settgast2023performant}.
The described approach aims to achieve high performance by introducing a data communication pattern on a single-level distributed memory architecture to support high-performance data movement from both cardinal and diagonal neighbors, as well as various optimizations to speed up overall performance.
This article provides both a dataflow implementation and evaluates the performance on a Cerebras \hbox{CS-2} machine.
We compare the performance with that of a GPU reference implementation evaluated on the latest NVIDIA GPUs.
This article makes the following contributions:

\begin{itemize}
\item it introduces a matrix-free  single-phase flow solver specifically designed for a dataflow architecture;
\item it presents a whole-fabric All-Reduce implementation on a dataflow architecture;
\item it describes an implementation of conjugate gradient algorithm on a dataflow architecture;
\item it describes algorithmic enhancements that accelerate performance on a dataflow architecture;
\item it presents a reference GPU implementation using the CUDA programming model; and
\item it evaluates performance on a Cerebras CS-2 machine and compares with a reference implementation on accelerated systems sporting NVIDIA GPUs, and demonstrates strong performance characteristics.
\end{itemize}

Section~\ref{sec:bg} introduces background and reviews related work.
Section~\ref{sec:mfcs2}  details a matrix-free FV kernel on a dataflow architecture.
Section~\ref{sec:mfcs2}  presents a GPU reference implementation for the same matrix-free FV kernel.
Section~\ref{sec:evaluation} evaluates performance and scalability and discusses the performance metrics using a Roofline~\cite{williams_roofline_2009} performance model.
Section~\ref{sec:concl} concludes and discusses future work.

\section{Background and Related Work}\label{sec:bg}

\subsection{Matrix-free Finite Volume Kernels for Geologic Carbon Capture and Storage (CCS) Simulation}\label{sec:geosx}

The simulation of flow and transport in geological porous media often relies on implicit finite-volume schemes \cite{lie2019introduction}. The most expensive step of the simulation workflow is the solution of the linear system that takes place at every time step (for linear problems) or at every Newton step (for nonlinear problem). Solving complex, ill-conditioned linear systems can account for at least 70\% of the computational time in realistic field-scale problems \cite{manea2015parallel}. This study leverages a dataflow architecture to accelerate a matrix-free solution strategy for the linear systems arising from a TPFA discretization of the equations governing single-phase flow in geological porous media. This is a key preliminary step towards adapting the complete set of discretized nonlinear multiphase flow equations to the dataflow model.

We consider an incompressible single-phase flow Darcy-scale model~\cite{aziz} able to represent the pressure changes due to fluid (supercritical CO$_2$) injection in a geological formation.
The system of governing equations consists of Darcy's law~\cite{darcy1856fontaines} and a mass balance equation:
\begin{subequations}
\begin{align}
& \boldsymbol{u} = - \frac{\kappa}{\mu} \nabla p, & & \mbox{(Darcy's law),} \label{eq:single_phase_darcy} \\
& \nabla \cdot \boldsymbol{u}  = 0, & & \mbox{(mass balance),} \label{eq:single_phase_mass_balance}
\end{align}
\end{subequations}
where $\boldsymbol{u}$ is the Darcy velocity, $\kappa$ is the (scalar) permeability coefficient, and $p$ is the pressure.
%
%
The viscosity, $\mu$, is assumed to be constant.
It is common to discretize this elliptic system by combining a low-order FV scheme with an implicit (backward-Euler) temporal discretization~\cite{aziz}.
The discretized system of nonlinear equations is:

\begin{equation}
  \mathbf{r}(\mathbf{p}) = \mathbf{0}, \label{eq:residual}
\end{equation}
where the residual $r_K = (\mathbf{r})_K$ in cell $K$ reads:
\begin{equation}
r_K =
\begin{cases}
    \sum_{L \in \text{adj}(K)} f_{KL}^{n+1}, & \text{if } K \notin \mathcal{T}^{\text{D}} \\[3pt]
    p_K - p_K^{\text{D}},              & \text{otherwise,}
\end{cases}
\label{eq:constrained_residual}
\end{equation}
in which the set $\mathcal{T}^{\text{D}}$ contains the cells where a Dirichlet boundary condition is imposed.
The interfacial flux, denoted by $f_{KL}^{n+1}$, where $L$ is an adjacent cell of $K$, is computed using a standard TPFA method as follows:

\begin{equation}
f^{n+1}_{KL} = \Upsilon_{KL} \lambda_{KL} (p^{n+1}_L - p^{n+1}_K). \label{eq:flux-calc}
\end{equation}
%
The transmissibility, $\Upsilon_{KL}$, is a coefficient accounting for the geometry of the cells and their permeability.
The (constant) interfacial fluid mobility, $\lambda_{KL}$, is computed as the arithmetic average of the mobilities in cells $K$ and $K$.

\begin{figure}[t]
  \centering
  \includegraphics[width=.45\linewidth, trim={0cm 0cm 0cm 0cm}, clip]{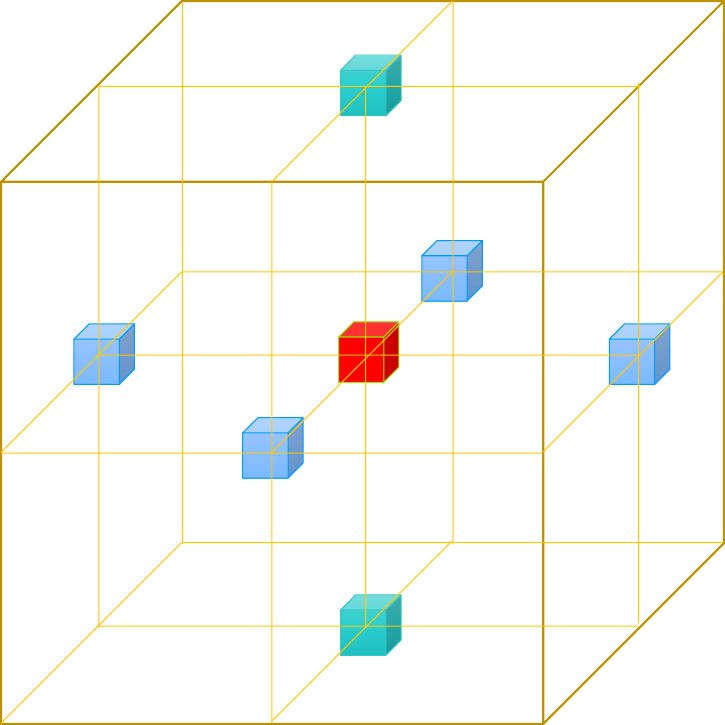}
  \caption{A 7-point stencil used in the flux computation.}\label{figure:tpfa-stencil}
\end{figure}

Performing a Newton step to update the pressure field $\mathbf{p}$ in Eq.~(\ref{eq:residual}) involves solving a linear system of the form:
\begin{equation}
  \mathbf{J} \delta \mathbf{p} = - \mathbf{r}, \quad \text{with} \quad \mathbf{J}_{ij}=\frac{\partial r_i}{\partial p_j}, \label{eq:jacobian}
\end{equation}
where the residual $\mathbf{r}$ is defined in Eq.~(\ref{eq:constrained_residual}).
The Jacobian matrix, $\mathbf{J}$, is symmetric positive definite, with a number of rows (and columns) equal to the number of cells in the mesh.
For large-scale systems, direct linear solvers are typically not practical and iterative Krylov-based methods are commonly employed instead.
We focus in this article on the conjugate gradient algorithm (Algorithm~\ref{alg:cg}).
This iterative Krylov-based solvers require the application of the Jacobian matrix $\mathbf{J}$ to a vector $\mathbf{x}$ to compute $\mathbf{J}\mathbf{x}$ at each linear iteration--see for example line  \ref{op1} in Algorithm~\ref{alg:cg}.
In a matrix-based approach, the full matrix $\mathbf{J}$ is assembled and stored in a sparse format, and then used in a second step to perform a standard matrix-vector multiplication.
In this article, we focus on the matrix-free approach, in which $\mathbf{J}$ is never fully assembled and stored.
Instead, local assembly and matrix-vector multiplication are fused by computing each entry of $\mathbf{J}\mathbf{x}$ as:
\begin{equation}
(\mathbf{J}\mathbf{x})_K =
\begin{cases}
    \displaystyle \sum_{L \in \text{adj}(K)} \Upsilon_{KL} \lambda_{KL} (x_L - x_K), & \text{if } K \notin \mathcal{T}^{\text{D}} \\[3pt]
    x_K              & \text{otherwise.}
\end{cases}
\label{eq:jx}
\end{equation}
The main advantages of the matrix-free approach are 1) to reduce the memory requirements by removing the need to store the full Jacobian matrix, and 2) to speedup the computations by removing the need to fill the global sparse Jacobian matrix.
Algorithm~\ref{alg:matrix-free} summarizes the operations involved in the matrix-free computation of $\mathbf{J}\mathbf{x}$.
The outer-loop of Algorithm~\ref{alg:matrix-free} sweeps over cells, and its inner-loop traverses a cell's neighbors.
For each neighbor, the contribution of the interface is computed and used to increment the corresponding entry of $\mathbf{J}\mathbf{x}$.
This study considers a 3D implementation of the algorithm on a Cartesian mesh, in which each interior cell has six neighbors (see Figure~\ref{figure:tpfa-stencil}).

\begin{algorithm}[t]
\caption{Conjugate gradient algorithm}
\label{alg:cg}
\begin{algorithmic}[1]
\State Compute $\mathbf{r}_0$ using Eq.~(\ref{eq:constrained_residual}).
\State $\mathbf{x}_0 \gets \mathbf{r}_0$
\State $k \gets 0$
\While{$(k < k_{\text{max}})$}
\State $\boldsymbol{\alpha}_k \gets \displaystyle{\frac{\mathbf{r}^T_k \mathbf{r}_k}{\mathbf{x}^T_k \mathbf{J} \mathbf{x}_k}}$ \label{op1}
\State $\mathbf{y}_{k+1} \gets \mathbf{y}_{k} + \boldsymbol{\alpha}_k \mathbf{x}_k$
\State $\mathbf{r}_{k+1} \gets \mathbf{r}_{k} - \boldsymbol{\alpha}_k \mathbf{J} \mathbf{x}_k$
\State if $\mathbf{r}^T_k \mathbf{r}_k < \epsilon$, then exit loop.
\State $\boldsymbol{\beta}_k  \gets \displaystyle{\frac{\mathbf{r}^T_{k+1} \mathbf{r}_{k+1}}{\mathbf{r}^T_k \mathbf{r}_k}}$
\State $\mathbf{x}_{k+1}  \gets \mathbf{r}_{k+1} + \boldsymbol{\beta}_k \mathbf{x}_k$
\State $k  \gets k+1$
\EndWhile
\end{algorithmic}
\end{algorithm}

\begin{algorithm}[t]
\caption{Matrix-free computation of $\mathbf{J}\mathbf{x}$ using Eq.~\ref{eq:jx}}
\label{alg:matrix-free}
\begin{algorithmic}[1]
\For{each cell $K$}
\If{$K \notin \mathcal{T}^{\text{D}}$}
\State $(\mathbf{J}\mathbf{x})_K \gets 0$
\For{each neighbor cell $L \in \text{adj}(K)$}
   \State $(\mathbf{J}\mathbf{x})_K \gets (\mathbf{J}\mathbf{x})_K + \Upsilon_{KL} \lambda_{KL} (x_L - x_K)$ \label{op2}
\EndFor
\Else
\State $(\mathbf{J}\mathbf{x})_K \gets x_K$
\EndIf
\EndFor
\end{algorithmic}
\end{algorithm}

\subsection{An Alternative Approach to Computing}
\label{sec:architecture}

One distinctive contribution of this work is the effective use of a dataflow architecture that is composed by a large number of processing elements.
In \cite{groq22b}, some of the same topics are introduced and discussed in detail.
In a current dataflow architecture main disruptive aspects are the independence from deep memory hierarchies, as opposed to all mainstream x86 architectures and GPGPUs, and the closeness of memory to the computing units, or processing elements. A common design achieving this will make use of a large wafer containing many instances of simple processing elements with its own memory interconnected via a low latency/high throughput fabric.

Algorithms to be ported to such architecture needs to be re-designed following at least two main aspects: data movements and re-use. A effective data distribution data is to key, in particular considering a large number of processing units with non-centralized coherent memory interface. Assuming this is achieved, next challenge is to efficiently re-used the data before sharing with neighboring computing units for the continuation of iterative algorithms. The latter concern is akin to what in traditional architecture would be handled with MPI or PGAS implementations. The former concern is similar to proper utilization of scratch-pad like memory, as opposed to traditional architectures, where the main challenge is to exploit the complex memory hierarchies. Also, most current HPC systems requires that developers are in full awareness of the memory hierarchy, on both, the host and the device sides of hybrid systems. In traditional architectures, the latter is usually addressed with CUDA/HIP, or highly portable programming models like RAJA/Kokkos or OpenMP.

\subsection{Accelerating Applications on Dataflow Architecture}

The utilization of dataflow architecture and its implementations for HPC workloads was initially explored in a previous study~\cite{DirkSC20}.
It focuses on a Python interface to address programming productivity issues related to low-level abstractions.
Our study is inspired by this work, but differs in our use of Cerebras Software Language (CSL) for implementation, which offers a higher level of abstraction and enables reusable functions and modules.
Additionally, our problem domain involves more complex computations and data communication patterns.
Further, our study focuses on achieving maximum performance on a real-world application that involves intricate data communication and requires computation with higher arithmetic intensity.

Jacquelin et al.~\cite{MauricioSC22} propose a memory layout for efficient computing of a large 3D data grid on a dataflow architecture, taking full advantage of its on-chip memory and 2D fabric.
Furthermore, a communication strategy is designed to exploit the architecture's fast communication fabric.
Computing the physics in our work demands more memory footprint and involves intricate data communication. This approach was fully exploited at the application level in~\cite{araya2023}.

Sai et al.~\cite{ryuichi_scalah_2023} explore a two-point flux approximation (TPFA) kernel for CCS on a dataflow architecture. In contrast, this paper describes the implementation of a full finite-volume kernel with a matrix-free approach for CCS on a dataflow architecture. 

Researchers have also used other dataflow-like architectures for scientific computation. Emani et al.~\cite{Samba21b} describe accelerating scientific applications on the Sambanova dataflow architecture. Also, Louw and McIntosh-Smith describe accelerating HPC computations including stencil-based Lattice-Boltzmann and a Gaussian filter on the Graphcore IPU~\cite{louw2021using}.

In a more general sense, dataflow or near-memory processing approaches have been highlighted as a potential path for post-exascale HPC~\cite{siam_tf_report}.

\section{Matrix-Free Finite Volume Kernels on a Dataflow Architecture}\label{sec:mfcs2}

This section details the implementation of a matrix-free solver for linear systems arising from a finite-volume (FV) discretization on a dataflow architecture.
First, we outline the mapping of both data (grid) and algorithm to the massively parallel mesh of processing elements (PE).
Next, we describe the data movement between cardinal PE neighbors needed for the FV computation for the horizontal plane.
Then, we present the all-reduce operator and the conjugate gradient algorithm deployed on this dataflow architecture.
Lastly, we describe the optimization techniques used to speed up performance.

\begin{figure*}[tb]
\centering
\begin{adjustbox}{width=.8\linewidth}
\includegraphics{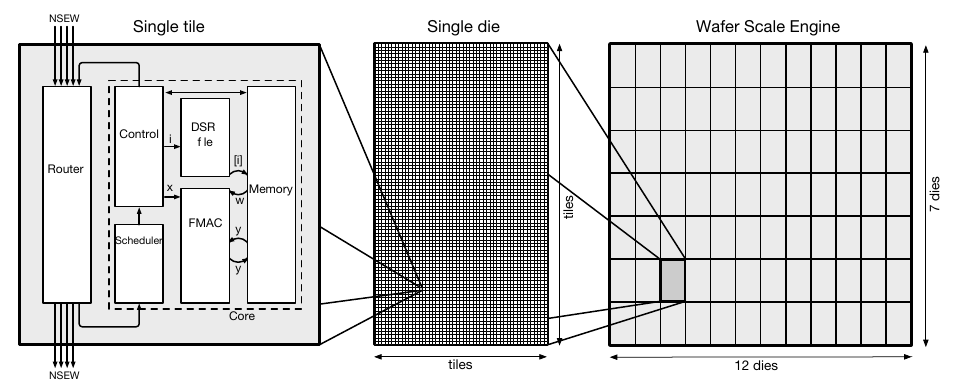}
\end{adjustbox}
\caption{
An overview of the Wafer Scale Engine (WSE). The \wse (to the right) occupies an entire wafer, and is a 2D array of dies. Each die is itself a grid of tiles (in the middle), which contains a processing element (to the left). Each processing element (PE) has its own router, that connects to the PE itself and the routes in its four cardinal neighboring routers. Each PE has 48KB of local memory and in total $\approx$ 850,000 PEs are available for computing. Figure credit:~\cite{MauricioSC22}
}
\label{figure:wse}
\end{figure*}

Figure~\ref{figure:wse} presents an overview of the Cerebras Wafer Scale Engine (WSE)~\cite{Cerebras-Micro23}. Like many dataflow architecture implementations, the WSE employs a 2D Cartesian mesh fabric to connect PEs. Each PE computes independently using data from its own private local memory. Each PE connects to a router to communicate with neighboring PEs. A PE's router manages five full-duplex links: a Ramp link that carries data between the PE and its router, while North, East, South, and West links connect a router to neighboring routers for inter-PE data communication. Links transfer data in 32-bit packets, each annotated with a color for routing and indicating the type of a message.

\subsection{Data Mapping}

\begin{figure}
  \centering
  \includegraphics[width=.54\linewidth]{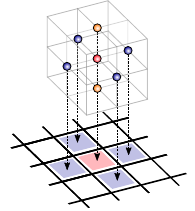}
  \caption{Three-dimensional problem mapping to a two-dimensional fabric of processing elements using a cell-based approach.}\label{figure:cell-based-mapping}
\end{figure}

In this study, the physical problem is represented by a 3D Cartesian mesh, where each cell has six neighbors.
On the X-Y plane (horizontal), each cell has four cardinal neighboring cells.
On the Z-dimension (depth), each cell only interacts with one cell above and cell one below.
Figure~\ref{figure:cell-based-mapping} illustrates the data mapping, it leverages the architecture's computation, memory, and communication resources.

Following the approach proposed by Jacquelin et al.~\cite{MauricioSC22},
We decompose the data domain such that every cell from the Z-dimension is mapped to the same PE, while the X and Y dimensions are mapped across the two axes of the fabric.
To be more precise, we map a cell with coordinates $(x, y, z)$ in the 3D mesh onto PE $(x, y)$.
A whole column of Z-dimension resides in the local memory of the corresponding PE.
This strategy, as demonstrated in prior works~\cite{DirkSC20,MauricioSC22}, optimizes the utilization of a dataflow architecture and maximizes the potential parallelism for HPC applications of this kind.

Each computation of Eq. (\ref{eq:jx}) requires three types of data:
the data from the current cell,
and the data exchanged between each pair of neighboring cells.
Figure~\ref{figure:cell-based-mapping} also depicts cell dependencies.
The center cell, along with its top and bottom neighbors, are mapped onto the PE colored in red.
Its cardinal neighboring cells and corresponding PEs are marked in purple.
Each PE allocates memory space for its current residual and pressure, as well as six transmissibilities for the computation of
Eq.~(\ref{eq:jx}) between the cell and its neighbors.
Each PE also allocates space to receive the pressure from all four neighboring cells on the same X-Y plane at each iteration.

\subsection{Distributed Data Communication in Matrix-Free Computation}\label{sec:data-comm}

Computing $(\mathbf{J}\mathbf{x})_K$ for each cell $K$ in Algorithm~\ref{alg:matrix-free} (line~\ref{op2}) requires data from six neighboring cells in all three dimensions.
Precisely, a cell $(x, y, z)$, depicted in red at the center of Figure~\ref{figure:tpfa-stencil}, requires
(a) cells from X-Y cardinal neighbors, as colored in blue in Figure~\ref{figure:tpfa-stencil}, corresponding to cells $(x+1,y,z)$,
 $(x-1,y,z)$,
 $(x,y+1,z)$, and
 $(x,y-1,z)$
 and
(b) cells $(x,y,z+1)$ and
 $(x,y,z-1)$ along the Z-dimension, yellow in Figure~\ref{figure:tpfa-stencil}. Given that each processing element PE already possesses data from its neighbors from the Z-dimension, no data movement is required.
Therefore, data communication happens only across the X-Y dimension of a 2D fabric.

\begin{table*}[t]
\centering
\begin{tabular}{|c|ccc|ccc|ccc|ccc|}
\hline
        & \multicolumn{3}{c|}{\begin{tabular}[c]{@{}c@{}}PEs with odd indices\\ on the X-dimension\end{tabular}}                            & \multicolumn{3}{c|}{\begin{tabular}[c]{@{}c@{}}PEs with even indices\\ on the X-dimension\end{tabular}}                            & \multicolumn{3}{c|}{\begin{tabular}[c]{@{}c@{}}PEs with odd indices\\ on the Y-dimension\end{tabular}}                            & \multicolumn{3}{c|}{\begin{tabular}[c]{@{}c@{}}PEs with even indices\\ on the Y-dimension\end{tabular}}                           \\ \hline
        & \multicolumn{1}{c|}{Action}                                                                 & \multicolumn{1}{c|}{AC} & CC  & \multicolumn{1}{c|}{Action}                                                                  & \multicolumn{1}{c|}{AC} & CC  & \multicolumn{1}{c|}{Action}                                                                 & \multicolumn{1}{c|}{AC} & CC  & \multicolumn{1}{c|}{Action}                                                                 & \multicolumn{1}{c|}{AC} & CC  \\ \hline
Step 1 & \multicolumn{1}{c|}{\begin{tabular}[c]{@{}c@{}}Send C \\ to East\end{tabular}}              & \multicolumn{1}{c|}{C1} & C5  & \multicolumn{1}{c|}{\begin{tabular}[c]{@{}c@{}}Receive \\ from West \\ into W\end{tabular}}  & \multicolumn{1}{c|}{C1} & C6  & \multicolumn{1}{c|}{\begin{tabular}[c]{@{}c@{}}Send C \\ to North\end{tabular}}             & \multicolumn{1}{c|}{C3} & C7  & \multicolumn{1}{c|}{\begin{tabular}[c]{@{}c@{}}Receive\\ from South\\ into S\end{tabular}}  & \multicolumn{1}{c|}{C3} & C8  \\ \hline
Step 2 & \multicolumn{1}{c|}{\begin{tabular}[c]{@{}c@{}}Receive \\ from West \\ into W\end{tabular}} & \multicolumn{1}{c|}{C2} & C6  & \multicolumn{1}{c|}{\begin{tabular}[c]{@{}c@{}}Send C \\ to East\end{tabular}}               & \multicolumn{1}{c|}{C2} & C5  & \multicolumn{1}{c|}{\begin{tabular}[c]{@{}c@{}}Receive\\ from South\\ into S\end{tabular}}  & \multicolumn{1}{c|}{C4} & C8  & \multicolumn{1}{c|}{\begin{tabular}[c]{@{}c@{}}Send C \\ to North\end{tabular}}             & \multicolumn{1}{c|}{C4} & C7  \\ \hline
Step 3 & \multicolumn{1}{c|}{\begin{tabular}[c]{@{}c@{}}Send C\\ to West\end{tabular}}               & \multicolumn{1}{c|}{C1} & C9  & \multicolumn{1}{c|}{\begin{tabular}[c]{@{}c@{}}Receive \\ from East \\ into E\end{tabular}}  & \multicolumn{1}{c|}{C1} & C10 & \multicolumn{1}{c|}{\begin{tabular}[c]{@{}c@{}}Send C \\ to South\end{tabular}}             & \multicolumn{1}{c|}{C3} & C11 & \multicolumn{1}{c|}{\begin{tabular}[c]{@{}c@{}}Receive\\ from North\\ into N\end{tabular}}  & \multicolumn{1}{c|}{C3} & C12 \\ \hline
Step 4 & \multicolumn{1}{c|}{\begin{tabular}[c]{@{}c@{}}Receive \\ from East \\ into E\end{tabular}} & \multicolumn{1}{c|}{C2} & C10 & \multicolumn{1}{c|}{\begin{tabular}[c]{@{}c@{}}Send C \\ to West\end{tabular}}               & \multicolumn{1}{c|}{C2} & C9  & \multicolumn{1}{c|}{\begin{tabular}[c]{@{}c@{}}Receive\\ from North\\ into N\end{tabular}}  & \multicolumn{1}{c|}{C4} & C12 & \multicolumn{1}{c|}{\begin{tabular}[c]{@{}c@{}}Send C \\ to South\end{tabular}}             & \multicolumn{1}{c|}{C4} & C11 \\ \hline
\multicolumn{6}{l}{AC: Action Color; CC: Completion Callback Color}
\end{tabular}
\caption{Data communication actions involved in each iteration.}
\label{table:tpfa-data-common-steps}
\end{table*}

Table~\ref{table:tpfa-data-common-steps} outlines the data communication actions for each iteration, alongside the associated colors for these actions and the colors designated for completion callbacks. Each row corresponds to one step in the data communication mechanism, totaling four steps. Upon completion of these steps, each cell contains data from all four neighboring cells. Each step involves four actions: for PEs with odd indices on the X-dimension, PEs with even indices on the X-dimension, PEs with odd indices on the Y-dimension, and PEs with even indices on the Y-dimension, respectively. These four actions within each step are executed concurrently, enabling the overlap of data communication and maximizing the bandwidth of the on-fabric network. After completing an action, such as sending or receiving all data, a designated callback color is triggered to notify the caller. The caller tracks the receipt of these completion callback colors. Although the four actions in each step can perform concurrently, progression to the next step only happens when all four completion callbacks are received, serving as a synchronization point.

In the X-Y plane, a PE communicates with its immediate neighbors along each cardinal direction of the fabric.
Specifically, for the X-dimension, a PE communicates with its eastbound neighbor at cell $(x+1,y,z)$ and its westbound neighbor at cell $(x-1,y,z)$.
For the Y-dimension, a PE communicates with its northbound neighbor at cell $(x,y-1,z)$ and its southbound neighbor at cell $(x,y+1,z)$.
Similar to the communication strategy proposed by Jacquelin et al.~\cite{MauricioSC22},
we dedicate two colors C$\{1,2\}$ for actions in X-dimension, and another two colors C$\{3,4\}$ for actions in Y-dimension.
Upon receiving the data, the corresponding flux computation will occur immediately in an event-driven fashion.
This strategy enables overlapping communications with computations.

During each communication step, multiple PEs simultaneously transmit data in all four directions. Each \textit{Sending} PE dispatches its local block of data in the respective direction. The transmitted data in each direction comprises arrays of \texttt{pressure}.

\begin{figure*}[t]
  \centering
  \begin{subfigure}[b]{2.7in}
      \centering
      \includegraphics[width=.35\textwidth]{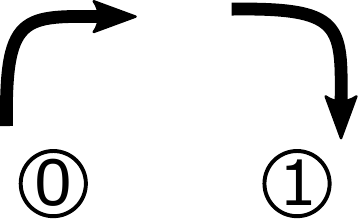}
      \caption{Router configurations used by FV flux computation. Configuration 0 corresponds to the switch position of a \textit{Sending} PE as the root of a broadcast, configuration 1 is the used by a \textit{Receiving} PE.}
      \label{fig:pe-router-configurations}
  \end{subfigure}
  \hfill
  \begin{subfigure}[b]{4.2in}
      \centering
      \includegraphics[width=.6\textwidth]{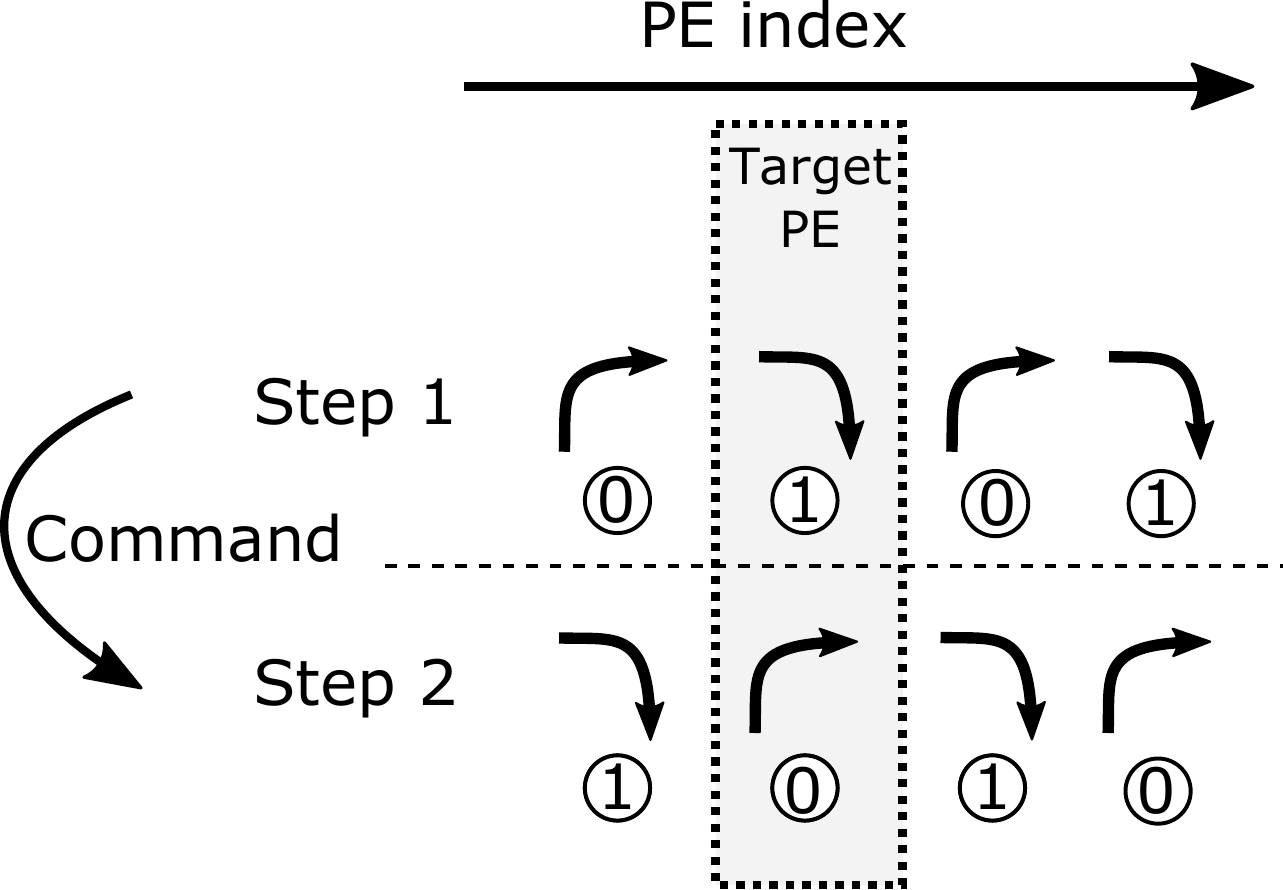}
      \caption{Two communication steps required to fetch all the data required by a target PE from the West. Corresponding router configurations are given in the circled numbers. At each step, a router command is sent through the broadcast pattern, changing the configurations from one to the alternative router configuration.}
      \label{fig:pe-communication-steps}
  \end{subfigure}
     \caption{Eastward localized broadcast operation used to exchange cell values along the X dimension.}
     \label{fig:pe-eastward-broadcast}
\end{figure*}

As shown in Figure~\ref{fig:pe-router-configurations}, two switch positions are programmed for each PE for sending and receiving accordingly.
Listing~\ref{lst:tpfa-eastward-broadcast} provides pseudo code illustrating this router configuration with the desired switch positions.

\begin{lstlisting}[float,floatplacement=t,language=C, basicstyle=\linespread{1.2}\ttfamily\scriptsize, label={lst:tpfa-eastward-broadcast}, caption=Pesudo CSL code for programming and controlling the router for an Eastward data communication as in Figure~\ref{fig:pe-eastward-broadcast}.]
  // PE Layout
  set_router_config(
    Cx, // Color used to advance switch position
    .{
      .routes = .{
        // First switch position
        .pos0 = { .rx = .{ RAMP }, .tx = .{ EAST } },

        // Second switch position
        .pos1 = { .rx = .{ WEST }, .tx = .{ RAMP } },

        // Ring mode on, allowing to revert to pos0
        // after pos1
        .ring_mode = true
      }
    }
  );

  // During PE execution, activate color Cx
  // when need to advance switch position
  mov32(fabric_control, .{ .activate = Cx });
  \end{lstlisting}

To switch a PE from a \textit{Sending} to a \textit{Receiving} state, we manage the PE router at runtime and alter its switch positions as needed. Once a \textit{Sending} PE sends its local data, it issues a command to change switch positions of its router and neighboring routers. In this process, the initial \textit{Sending} PE transitions to a \textit{Receiving} PE, while the original \textit{Receiving} PE becomes a \textit{Sending} one. Listing~\ref{lst:tpfa-eastward-broadcast} provides pseudo-code depicting this switch adjustment process. Subsequently, the new \textit{Sending} PE transmits its local data and switch command to itself and its neighbors. This pattern of alternating switch positions continues throughout the entire data communications. Since data and commands are exchanged only with immediate neighbors, the PE seamlessly toggles between \textit{Sending} and \textit{Receiving} states, as depicted in Figure~\ref{fig:pe-communication-steps}. After two steps, all PEs have sent and received the required data.

\subsection{Whole-Fabric All-Reduce}\label{sec:all-reduce}

Computing $\boldsymbol{\alpha}_k$ (line 5) and $\boldsymbol{\beta}_k$ (line 9) of the Algorithm~\ref{alg:cg} requires dot product calculations across all PEs on the 2D fabric. Therefore, we derived an implementation of the \verb|all-reduce| operation on a dataflow system. It follows a three-step algorithm:
\begin{enumerate}

\item Horizontal reductions for all the rows are performed, where the reduction occurs from left to right of every row. The right-most PE in each row receives all values from the same row. The updated result for each row is stored in the right-most PE for the respective row, while other PEs do not change the content.
\item A right-most column reduction is conducted, moving from top to bottom. The bottom-right PE collects all values and computes the final result. The bottom-right updates itself with the final result.
\item The bottom-right PE broadcasts the data upwards to the entire right-most column. Then, each PE in this column broadcasts data to all of its left PEs to cover the entire fabric. When broadcasted to a PE, the PE updates its content. Upon the completion of the broadcast, all PEs hold the updated value.
\end{enumerate}

This is implemented as an asynchronous task, and when the process finishes, it triggers a callback task to continue the rest of the program execution.

\subsection{Conjugate Gradient Implementation}

The conjugate gradient (CG) algorithm is an iterative approach, where it continuously iterates the search process until an acceptable approximation to the solution is obtained. On conventional architectures, its implementation usually involves nested loops, and once the specified tolerance condition is satisfied, the algorithm exits the main loop.

Unlike the conventional approach, our implementation of the conjugate gradient algorithm on a dataflow architecture utilizes a state machine (a previous attempt by \cite{DirkSC20} implemented CG on a WSE using a different approach and programming language). We have devised 14 states to orchestrate the various steps involved in the conjugate gradient algorithm (Algorithm~\ref{alg:cg}) and have carefully planned the transitions between these states. Below, we outline some of the key states and their transitions.

The \verb|INIT| state sets up the state machine and initializes the residual and search direction (referred to as lines 1~and~2 in Algorithm~\ref{alg:cg}).

The \verb|ITER_CHECK| and \verb|THRES_CHECK| states handle iteration checking (line~4 in Algorithm~\ref{alg:cg}) and convergence checking (line 8 in Algorithm~\ref{alg:cg}), respectively. All conditional checks, whether for \verb|if| or \verb|while| statements, are converted into state transitions.

In Section~\ref{sec:data-comm}, we discuss the data communication required for computing $(\mathbf{J}\mathbf{x})_K$ on the right-hand side of line~5 in Algorithm~\ref{alg:cg}. Once all neighboring cells are received, it executes the matrix-free FV computation.

Section~\ref{sec:all-reduce} demonstrates the use of the \verb|all-reduce| operator in computing the remaining portions of the right-hand side of line 5 and line 9 in Algorithm~\ref{alg:cg}, respectively.

\subsection{Algorithmic enhancements}

We highlight some of the most effective and impactful algorithmic improvements in terms of memory utilization and computing performance.
First, a brief reminder, the cells in the same vertical column share the private memory of a PE, therefore reducing the memory consumption on each PE is crucial to fit the largest possible computational domain.
With respect to performance, asynchronous communications between PEs are necessary to overlap data movement with computations. Further, vectorization is used to exploit the bandwidth of the fabric.

\subsubsection{PE private memory saving strategies}

Within its local private memory, each PE must retain instructions and all necessary data for computational tasks. Besides storing cell and face data, the local memory must accommodate extra buffers for both data broadcasting and computations. Each PE has only 48 KiB memory space, making the reuse of data buffers important.
This approach is analogous to register allocation optimization, which is typically managed by compilers but is manually handled in our implementation. 
The impact of this optimization is twofold: first, larger simulations can be tackled by minimizing the implementation's memory footprint. Second, in certain scenarios, overwriting or reusing data buffers eliminates the necessity for data replication, thereby enhancing computational efficiency.

\subsubsection{Asynchronous communications}

The implementation of the proposed algorithm relies on asynchronous communications to exchange data among PEs, occurring asynchronously along four directions. This approach has two advantages. Firstly, simultaneous transfers enhance the potential for leveraging interconnect-level concurrency. Secondly, non-blocking communications enable the overlapping of transfers with useful computations, effectively hiding associated overheads to the maximum extent possible. This is possible because the fabric and routers operate entirely independently from the PEs.

\subsubsection{Vectorization of floating-point operations}

In many hardware architectures, dedicated mechanisms exist for processing arrays of data, often referred to as vectors. In the Cerebras architecture, this functionality is achieved through special registers known as \textit{Data Structure Descriptors (DSDs)}, which serve as vectors upon which specific instructions can operate. The DSDs contain information regarding the address, length, and stride of the arrays on which a given instruction can operate upon. It enables vectorized instructions to fully utilize Single Instruction, Multiple Data (SIMD) units, with up to two units available for single precision arithmetic at the hardware level. From a dataflow perspective, an instruction acting on DSDs functions as a filter through which data flows. Notably, regardless of the size of the input and output arrays (within memory constraints), the instruction's throughput remains constant since there is no involvement of caching. Consequently, the utilization of these dedicated registers significantly enhances performance.

\section{Matrix-Free Finite Volume Kernels on a GPU}\label{sec:mfgpu}

This section describes our reference matrix-free finite volume (FV) implementation on GPUs as a baseline compared to the one on implemented a dataflow architecture. The reference implementation employs a standard implicit FV flux calculation based on TPFA. The FV numerical method is similar to those adopted by commercial simulators and benchmark open-source simulators such as GEOS~\cite{GEOSXWebsite}.

To fairly compare to our implementation on a dataflow architecture, our reference implementation also follows a matrix-free approach. To begin with, memory is allocated on both host and device memory. Then, the mesh of size $N_x \times N_y \times N_z$ is loaded into host memory, with the X-dimension as the innermost dimension and Z-dimension as the outermost dimension in the memory layout.
Next, we copy all data from host to device memory, which is overall the same procedure followed with the CS2, fully occupying the device and not using the host at all.
Since we evaluate our GPU kernel on the latest hardware with large enough device memory to load all data at once,
we avoid data domain decomposition and avoid frequent data transfers between host and device memory.

For each time iteration of the simulation, we launch a GPU kernel with 3D threadblocks, carefully mapping each cell to a GPU thread.
Our kernel launch is implemented using CUDA programming model.
We launch 3D GPU threadblocks, each with a size of $1024$ to respect the GPU's limit of at most $1024$ threads per block, while maximizing the thread parallelism and GPU utilization.
The threadblock on a nested loop space of the whole data mesh size of $N_x \times N_y \times N_z$, we launch GPU threadblock size of $16\times 8 \times 8$, where $16$ is the innermost dimension size.

Each GPU kernel is scheduled to concurrently invoke a device function that performs the matrix-free FV computation following Algorithms~\ref{alg:cg}~and~\ref{alg:matrix-free}.
In a nutshell, the outer loop (line 4 in Algorithm~\ref{alg:cg}) implements the conjugate gradient iterative method  Algorithm~\ref{alg:cg}.
Next, the for loop on line 1 in Algorithm~\ref{alg:matrix-free} is offloaded to GPU, and each GPU thread handles a cell $K$.
Then, the inner loop (line 4 in Algorithm~\ref{alg:matrix-free}) performs the matrix-free FV computation, as described in Algorithm~\ref{alg:matrix-free}.
Each thread concurrently fetches the cell data for itself and all cell data from its six neighboring cells.
For each neighbor, it computes Eq.~(\ref{eq:jx}) using the transmissibility, the local cell values, its neighboring values, and produces a local interfacial contribution to $(\mathbf{J}\mathbf{x})_K$.
Finally, it assembles all the local fluxes and updates the current cell value.

\section{Experimental Evaluation}\label{sec:evaluation}

We evaluate our dataflow implementation on a Cerebras \hbox{CS-2} system and compare it to the reference implementation running on NVIDIA GPUs.
We discuss their respective performance characteristics.

\begin{figure*}
    \centering
    \includegraphics[scale=0.5]{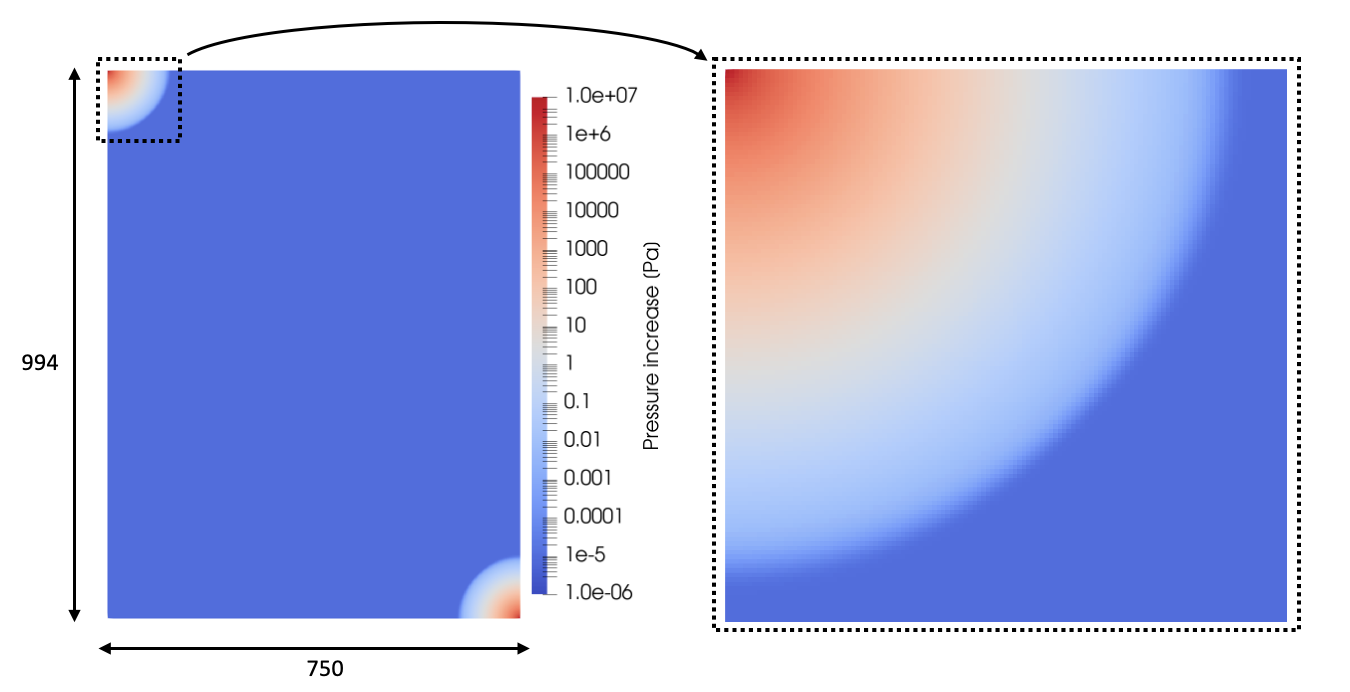}
    \caption{Pressure propagation from the source (top left in the left plot) to the producer (bottom right of the mentioned plot). The right plot depicts the source in detail.}
    \label{fig:pressure}
\end{figure*}

\subsection{Experimental Configurations}

Our experiments use the CS2 fabric in its entirely, therefore the grid size is $750 \times 994$.
This is the maximum size allowed by the SDK as a thin layer of PEs around the boundary of the fabric is reserved for system use.
While the CS-2 is connected to a Linux host, the server is only used to schedule the workload, and no computations take place on the Linux host during the experiments.
The software stack is composed of the recently released Cerebras SDK 1.0.0, which includes a faster compiler, additional utility libraries, such as \textit{memcpy},
and offers improved performance compared to the previous version reported for instance in~\cite{MauricioSC22}.

A100 and H100-based platforms were used to obtain the reference experiments.
The host of the A100 system has a 16-core AMD EPYC 7F52 CPU with 256 GB of RAM.
The computing node is also equipped with four NVIDIA A100 GPUs, each has 40 GB of on-device memory.
Our experiments only use one of the A100 GPUs sported by the computing node.
Our toolchain for the A100-based experiments consists of the 
NVIDIA driver version \hbox{530.30.02},
CUDA 12.1, and GCC 8.3.1.
The H100 GPU results included in this study were obtained from the H100, which is part of a Grace Hopper superchip. It sports 16,896 CUDA cores and 95 GB of device memory.
Its NVIDIA driver version is \hbox{535.154.05} with CUDA 12.2 as part of NVHPC release 23.11.
We compare and numerically validate the CS-2 results to the reference implementations.

\subsection{Numerical Integrity}

We compare and validate the numerical results produced by the CS-2 to those yielded by the reference implementation running on GPUs.

In addition, we plot the results (see Figure~\ref{fig:pressure}), showing the final state of the pressure in the grid after the algorithm has converged.

\subsection{Performance Results}

In this section, performance results are provided for matrix-free FV computation.
We discuss various observations to provide insights on WSE-2's performance characteristics.

\paragraph{Measurements and comparisons}

Table~\ref{tbl:time-measurements} presents time measurements, each applying the FV kernel to a $750 \times 994 \times 922$ mesh. The simulation converges in 225 iterations for a tolerance of ${2 \times {10^{-10}}}$.
The performance of our dataflow implementation is shown in the Dataflow/CSL row.
We present the reference implementation running on an NVIDIA A100 GPU and H100 GPU in the A100/CUDA and H100/CUDA rows, respectively.
We report the average kernel time and standard deviation for each implementation from multiple runs. For all architectures, all floating-point numbers used in the experiments are 32-bit.

\begin{table}
\centering
\begin{tabular}{|c|r|l|}
\hline
\multicolumn{1}{|c|}{Arch/lang} & \multicolumn{1}{c|}{Avg.} & \multicolumn{1}{c|}{S.D.} \\ \hline
Dataflow/CSL                   & 0.0542                    & 0.000014                 \\ \hline
A100/CUDA              & 23.1879                   & 0.123267                 \\ \hline
H100/CUDA              & 11.3861                   & 0.222566                 \\ \hline
\end{tabular}
\caption{Time measurements on CS-2 and NVIDIA GPUs.}
\label{tbl:time-measurements}
\end{table}

All of these show that our reference implementations have high performance.
Our dataflow implementation, compared to the GPU reference implementation running on H100 and A100, achieves a speedup of 209.68x and 427.82x, respectively.

\paragraph{Scaling comparison}

Inspired by the experiment as shown in~\cite{MauricioSC22}, we also conduct a weak scaling experiment. Our experiment consists of two parts:
The first one focuses solely on the computation outlined in Algorithm~\ref{alg:matrix-free}, aimed at demonstrating the scalability of a matrix-free FV kernel across the PE dimensions on the CS-2.
The second phase involves the entire execution as depicted in Algorithm~\ref{alg:cg}, showcasing the scalability of the entire matrix-free FV approach.
We modify the grid dimension along the X and Y dimensions while keeping the Z-dimension constant.
The X and Y dimensions are stretched up to the maximum size of $750 \times 994$.

On CS-2, time is recorded via the SDK's \verb|<time>| library, which uses hardware timestamp counters.
On A100, we use CUDA provided runtime functions, such as \textit{cudaEventRecord}, \textit{cudaEventSynchronize}, and \textit{cudaEventElapsedTime}, to measure the elapsed time for GPU kernel runs.

Table~\ref{tbl:weak-scaling} shows the grid size of each experiment and the number of time steps required to converge.
Then, two sections show the scalability of Algorithms~\ref{alg:matrix-free}~and~\ref{alg:cg}, respectively.
Within each section, we show the throughput achieved on CS-2 in Gigacell/s, the wall-clock time required to convergence given the grid size, as well as the kernel time of our CUDA reference implementation running on A100.
Timings reported here correspond to computations taking place on the device only.

\begin{table*}
\centering
\begin{tabular}{|l|l|l|r|r|||r|r||r|||r|r||r|}
\hline
\multicolumn{5}{|c|||}{Grid} & \multicolumn{3}{c|||}{Algorithm~\ref{alg:matrix-free}} & \multicolumn{3}{c|}{Algorithm~\ref{alg:cg}} \\ \hline
\multicolumn{1}{|c|}{Nx} & \multicolumn{1}{c|}{Ny} & \multicolumn{1}{c|}{Nz} & \multicolumn{1}{c|}{\begin{tabular}[c]{@{}c@{}}Total Number \\ of Cells\end{tabular}} & \multicolumn{1}{c|||}{\begin{tabular}[c]{@{}c@{}}Num. of Steps \\ to Converge\end{tabular}} & \multicolumn{1}{c|}{\begin{tabular}[c]{@{}c@{}}Throughput \\ {[}Gcell/s{]}\end{tabular}} & \multicolumn{1}{c||}{\begin{tabular}[c]{@{}c@{}}CS-2\\ time {[}s{]}\end{tabular}} & \multicolumn{1}{c|||}{\begin{tabular}[c]{@{}c@{}}A100\\ time {[}s{]}\end{tabular}} & \multicolumn{1}{c|}{\begin{tabular}[c]{@{}c@{}}Throughput \\ {[}Gcell/s{]}\end{tabular}} & \multicolumn{1}{c||}{\begin{tabular}[c]{@{}c@{}}CS-2\\ time {[}s{]}\end{tabular}} & \multicolumn{1}{c|}{\begin{tabular}[c]{@{}c@{}}A100\\ time {[}s{]}\end{tabular}} \\ \hline
200  & 200 & 922 & 36,880,000                & 226 & 680.43    & 0.0122 & 1.3979               & 330.79  & 0.0251 & 2.8021             \\ \hline
400  & 400 & 922 & 147,520,000               & 225 & 2,721.57  & 0.0122 & 2.7743               & 982.72  & 0.0337 & 5.6343             \\ \hline
600  & 600 & 922 & 331,920,000               & 225 & 6,122.27  & 0.0122 & 5.2882              & 1,764.34 & 0.0423 & 11.8380            \\ \hline
750  & 600 & 922 & 414,900,000               & 225 & 7,653.38  & 0.0122 & 7.1703              & 2,044.08 & 0.0456 & 16.3473            \\ \hline
750  & 800 & 922 & 553,200,000               & 225 & 10,204.11 & 0.0122 & 9.1577              & 2,487.70 & 0.0500 & 20.9367            \\ \hline
750  & 950 & 922 & 656,925,000               & 225 & 12,115.52 & 0.0122 & 9.2548              & 2,776.97 & 0.0532 & 22.9128            \\ \hline
750  & 994 & 922 & 687,351,000               & 225 & 12,688.55 & 0.0122 & 9.5507              & 2,855.48 & 0.0542 & 23.1879            \\ \hline
\end{tabular}
\caption{Results of various grid sizes for scalability experiments. The Throughput column is computed only for the CS2.}
\label{tbl:weak-scaling}
\end{table*}

Table~\ref{tbl:weak-scaling} illustrates that Algorithm~\ref{alg:matrix-free} exhibits nearly perfect scaling on WSE-2 across the PE dimensions.
The results also indicate that the execution time on WSE-2 increases for Algorithm~\ref{alg:cg} as both the X and Y dimensions expand.
This observed increase in time can be attributed to the involvement of reduction operators in the conjugate gradient algorithm.
As the 2D fabric expands, more values need to be computed by the reduction operator, and data also needs to travel longer distances across the fabric.
Nevertheless, our data reveals that for both algorithms, the scalability of WSE-2 surpasses that of a GPU.

\paragraph{Data communication cost}

To gain further insight into the performance characteristics of the matrix-free FV algorithm on the CS-2, we conducted an experiment wherein we modified our dataflow implementation to exclude all floating-point operations, focusing solely on measuring the time required for data communications. This modified version of the code was executed on a mesh of size $750 \times 994 \times 922$, maintaining consistency with the grid size used in our time measurements. However, as the run without computation never converged, we terminated it at step 225 to align with our time measurement experiment.

From this experiment, we obtained an average data communication time of $0.0034$ seconds. Considering a total device time of $0.0542$ seconds, we can infer that the computation consumes $0.0508 \sim 0.0542$ seconds, with $0.0508$ representing the non-overlapped time for floating-point calculations. This duration could potentially take as much as the whole execution time if communication were perfectly overlapped with computation.

\begin{table}
\centering
\begin{tabular}{|c|r|r|}
\hline
              & \multicolumn{1}{c|}{Time {[}s{]}} & \multicolumn{1}{c|}{Percentage {[}\%{]}} \\ \hline
Data Movement & 0.0034                            & 6.27                                    \\ \hline
Computation   & 0.0508 $\sim$ 0.0542              & 93.73 $\sim$ 100.00                      \\ \hline
Total         & 0.0542                            & 100.00                                   \\ \hline
\end{tabular}
\caption{Time distribution on CS-2 with our largest possible mesh size.}
\label{tbl:time-distribution}
\end{table}

Table~\ref{tbl:time-distribution} presents the results of our time distribution analysis.
We observed that the data communication cost accounted for $6.27\%$ of the total device time, while the remaining $93.72\%$ was dedicated to non-overlapping floating-point computations with computation could potentially take up to $100\%$ of the whole execution time in a perfectly overlapped fashion.
This analysis provides a more detailed understanding of the performance characteristics of the WSE-2 for the matrix-free FV kernel, which can be useful for optimizing future implementations of this algorithm on this platform.

\subsection{Roofline}

The Roofline~\cite{williams_roofline_2009, yang_hierarchical_roofline_2020} model provides a visual representation of a code's performance with relative to a machine's peak performance.
This model takes into account the code's arithmetic intensity, memory bandwidth, and overall performance and presents them on a single chart.
By comparing a code's performance to the platform's performance ceiling, the Roofline model can also provide valuable insights into potential optimizations for the code.

To compute the performance characteristics of our dataflow implementation on the roofline chart,
we took the same roofline model described in~\cite{MauricioSC22}.
The summary of instructions at the cell level needed for the matrix-free FV algorithm,
including their floating-point operation count, memory traffic, and fabric traffic, are detailed in Table~\ref{tbl:wse-operations}.
Each PE performs computations for $N_z$ cells, with each cell performing a matrix-free computation with six neighbors.
Computing line  \ref{op2} in Algorithm~\ref{alg:matrix-free} consists of 6 FMULs, 4 FSUBs, 1 FADD, 1 FMA, and 1 FNEG, with FMA requiring two FLOPs and other operations requiring one FLOP.
Therefore, computing with one neighbor requires 14 FLOPs, and each cell computing with all six neighbors performs a total of 84 FLOPs.
The rest of the computations in Algorithm~\ref{alg:cg} perform 2 FMULs and 5 FMAs, totaling 12 FLOPs.
In total, each cell computing the matrix-free FV kernel performs a total of 96 FLOPS.

\begin{table*}
\centering
\begin{tabular}{|c||c|c||c|c|c|}
\hline
Area & Operation & Counts & FLOP & Memory traffic     & Fabric traffic \\ \hline
\multirow{6}{*}{Alg.~\ref{alg:matrix-free}} & FMUL      & 36     & 1    & 2 loads, 1 store & 0              \\ \cline{2-6}
 & FSUB      & 24     & 1    & 2 loads, 1 store & 0              \\ \cline{2-6}
 & FNEG      & 6     & 1    & 1 load, 1 store  & 0              \\ \cline{2-6}
 & FADD      & 6     & 1    & 2 loads, 1 store & 0              \\ \cline{2-6}
 & FMA       & 6     & 2    & 3 loads, 1 store & 0              \\ \cline{2-6}

 & FMOV      & 4     & 0    & 1 store          & 1 load         \\ \hline

\multirow{3}{*}{Rest of Alg.~\ref{alg:cg}} & FMUL      & 2     & 1    & 2 loads, 1 store & 0              \\ \cline{2-6} 
 & FMA       & 5     & 2    & 3 loads, 1 store & 0              \\ \cline{2-6} 
 & FMOV      & 4     & 0    & 1 store          & 1 load         \\ \hline 
\end{tabular}
\caption{Instruction and memory access counts for one mesh cell on CS-2. }
\label{tbl:wse-operations}
\end{table*}





\begin{figure}
  \centering
  \includegraphics[width=0.9\linewidth, trim={9cm 1.5cm 7cm 1.5}, clip]{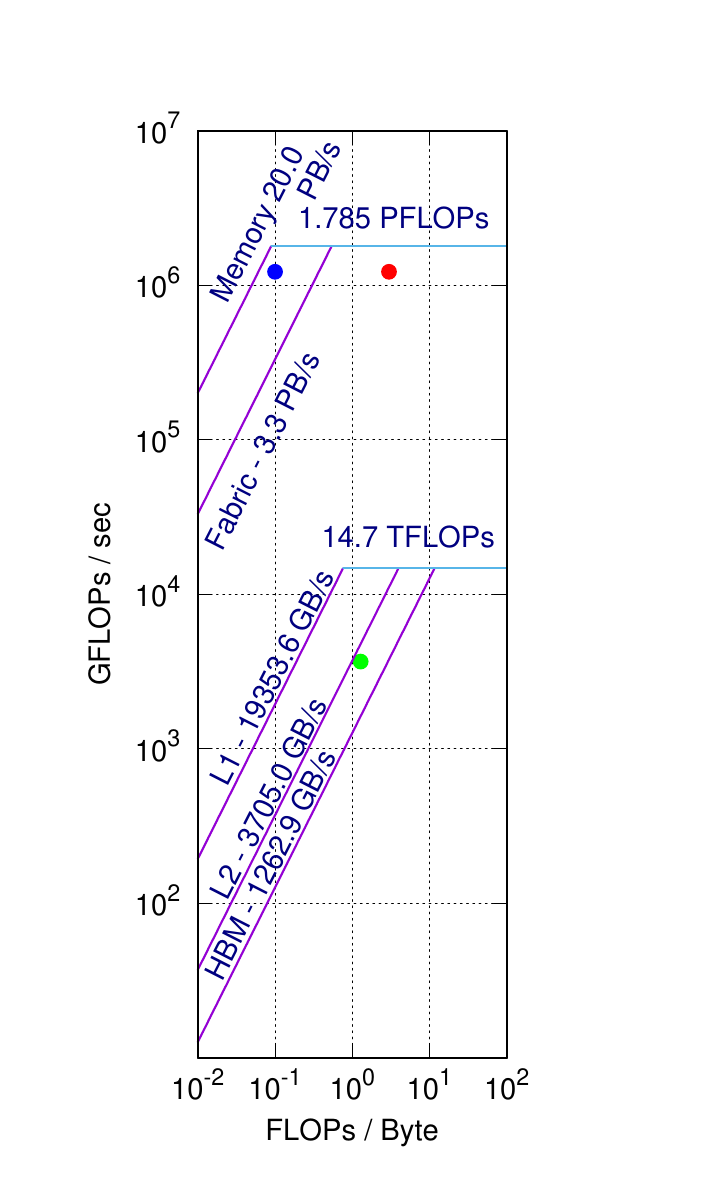}
  \caption{Roofline models for CS-2 and A100 (in log-log scale) for a $750 \times 994 \times 922$ mesh. Dots represent the implementations of matrix FV kernel.
 For CS-2 (top), there are two distinct resources - memory and fabric. The blue dot on the left represents arithmetic intensity with respect to memory accesses, while the red dot on the right corresponds to arithmetic intensity with respect to fabric accesses. The kernel is compute-bound for both memory accesses and fabric access, achieving 68\% of peak performance.
 A100 (bottom) is depicted in green. The kernel is memory-bound, achieving 78\% of the peak performance. }
  \label{figure:tpfa-roofline}
\end{figure}

The floating-point operations perform a total of 268 loads and stores of single-precision 32-bit floating-point numbers to and from memory, and 8 loads from fabric.
Four of these fabric loads represent the data communications with the four neighboring PEs, while the other four fabric loads are for a row reduction, a column reduction, a row broadcast, and a column broadcast, respectively.
Data accesses from top and bottom cells in the mesh only require memory access since they are in the same PE's memory and do not require fabric accesses.
As a result, the arithmetic intensity is 0.0895 FLOPs/Byte with respect to memory access and 3 FLOPs/Byte with respect to fabric transfers.
With a mesh size of $750 \times 994 \times 922$ and the computation being completed in 0.0542s, it achieves a performance of 1.217~PFLOPS, as depicted in Figure~\ref{figure:tpfa-roofline}~(top).
Our dataflow implementation is compute-bound for both memory access and fabric access.
They both achieve $68.18\%$ of machine peak performance.
This provides insights for future performance improvements.


In the GPU case, we measure our machine performance characteristics using Empirical Roofline Toolkit~\cite{yang_roofline_ert_2020} and characterize the reference implementation using NVIDIA's Nsight Compute on A100.
We present GPU roofline in Figure~\ref{figure:tpfa-roofline}~(bottom). The roofline plot validates the efficiency of our GPU implementation, therefore, comparisons against it are sound. 

\section{Conclusions and Future Work}\label{sec:concl}

This study presents a highly efficient matrix-free FV algorithm implementation that takes advantage of low-latency localized communications and single-level memory architecture.
We present distributed data communications and a \verb|all-reduce| process on a dataflow architecture.
We describe a conjugate gradient implementation on a dataflow architecture.
We demonstrate memory-saving optimizations by reusing space for intermediate results.
We describe our strategies to speed up computations with vectorization of floating-point operations and overlapping data movements with computations.

Our experiments show that our dataflow implementation achieves 1.217 PFLOPS on a single-node CS-2 machine and
a significant speedup of 427x compared to a reference implementation running on an NVIDIA A100 GPU.
The matrix-free FV algorithm presents strong weak scaling in the fabric dimensions on a dataflow architecture.
We show additional performance characteristics for the matrix-free FV algorithm on a system using the dataflow architecture and demonstrate it can be exploited efficiently for this kind of application with massive throughout.

Future work includes supporting arbitrary mesh topologies and mapping them efficiently onto a dataflow architecture to enable porting of a broader range of FV applications.
We also need to develop data broadcasting strategies to support data movement from any cell in the arbitrary-shaped mesh.

\section*{Acknowledgment}

The authors would like to thank TotalEnergies for allowing us to share the material.
We thank Shelton Ma and Harry Zong from TotalEnergies for all the support.
We wish to acknowledge Mathias Jacquelin, Leighton Wilson, and Michael James from Cerebras for the introductions and instructions, especially for Mathias for contributing to an earlier version of this manuscript.
We recognize Randolph Settgast from Lawrence Livermore National Laboratory for all assists.

\bibliographystyle{IEEEtran}
\bibliography{main}

\begin{thebibliography}{10}
\providecommand{\url}[1]{#1}
\csname url@samestyle\endcsname
\providecommand{\newblock}{\relax}
\providecommand{\bibinfo}[2]{#2}
\providecommand{\BIBentrySTDinterwordspacing}{\spaceskip=0pt\relax}
\providecommand{\BIBentryALTinterwordstretchfactor}{4}
\providecommand{\BIBentryALTinterwordspacing}{\spaceskip=\fontdimen2\font plus
\BIBentryALTinterwordstretchfactor\fontdimen3\font minus \fontdimen4\font\relax}
\providecommand{\BIBforeignlanguage}[2]{{%
\expandafter\ifx\csname l@#1\endcsname\relax
\typeout{** WARNING: IEEEtran.bst: No hyphenation pattern has been}%
\typeout{** loaded for the language `#1'. Using the pattern for}%
\typeout{** the default language instead.}%
\else
\language=\csname l@#1\endcsname
\fi
#2}}
\providecommand{\BIBdecl}{\relax}
\BIBdecl

\bibitem{srccs_report}
{Intergovernmental Panel on Climate Change}, \emph{{Carbon Dioxide Capture and Storage}}.\hskip 1em plus 0.5em minus 0.4em\relax Cambridge University Press, 2005.

\bibitem{reed2022reinventing}
D.~Reed, D.~Gannon, and J.~Dongarra, ``{Reinventing High Performance Computing: Challenges and Opportunities},'' \emph{arXiv:2203.02544}, 2022.

\bibitem{MauricioIBMCellBE}
\BIBentryALTinterwordspacing
M.~Araya-Polo, F.~Rubio, R.~de~la Cruz, M.~Hanzich, J.~M. Cela, and D.~P. Scarpazza, ``{3D Seismic Imaging through Reverse-Time Migration on Homogeneous and Heterogeneous Multi-Core Processors},'' \emph{Sci. Program.}, vol.~17, no. 1–2, p. 185–198, January 2009. [Online]. Available: \url{https://doi.org/10.1155/2009/382638}
\BIBentrySTDinterwordspacing

\bibitem{kahle1989connection}
B.~A. Kahle and W.~D. Hillis, ``{The Connection Machine model CM-1 architecture},'' \emph{IEEE Transactions on Systems, Man, and Cybernetics}, vol.~19, no.~4, pp. 707--713, 1989.

\bibitem{cerebras_hotchips2019}
\BIBentryALTinterwordspacing
Cerebras, ``{Wafer-Scale Deep Learning},'' in \emph{2019 IEEE Hot Chips 31 Symposium (HCS)}.\hskip 1em plus 0.5em minus 0.4em\relax Los Alamitos, CA, USA: IEEE Computer Society, August 2019, pp. 1--31. [Online]. Available: \url{https://doi.org/10.1109/HOTCHIPS.2019.8875628}
\BIBentrySTDinterwordspacing

\bibitem{groq_HotChip}
D.~Abts, J.~Kim, G.~Kimmell, M.~Boyd, K.~Kang, S.~Parmar, A.~Ling, A.~Bitar, I.~Ahmed, and J.~Ross, ``{The Groq Software-defined Scale-out Tensor Streaming Multiprocessor : From chips-to-systems architectural overview},'' in \emph{{2022 IEEE Hot Chips 34 Symposium (HCS)}}, 2022, pp. 1--69.

\bibitem{Samba21}
R.~Prabhakar and S.~Jairath, ``{SambaNova SN10 RDU:Accelerating Software 2.0 with Dataflow},'' in \emph{{2021 IEEE Hot Chips 33 Symposium (HCS)}}.\hskip 1em plus 0.5em minus 0.4em\relax IEEE, 2021, pp. 1--37.

\bibitem{GEOSXWebsite}
\BIBentryALTinterwordspacing
L.~L.~N. Laboratory, S.~University, TotalEnergies, and G.~Contributors, ``{GEOS: Next-gen simulation for geologic carbon storage},'' \textsc{url:}~\url{https://www.geos.dev/}, Mar. 2023. [Online]. Available: \url{https://www.geos.dev/}
\BIBentrySTDinterwordspacing

\bibitem{settgast2023performant}
R.~R. Settgast, Y.~Dudouit, N.~Castelletto, W.~R. Tobin, B.~C. Corbett, and S.~Klevtsov, ``Performant low-order matrix-free finite element kernels on gpu architectures,'' \emph{arXiv preprint arXiv:2308.09839}, 2023.

\bibitem{williams_roofline_2009}
\BIBentryALTinterwordspacing
S.~Williams, A.~Waterman, and D.~Patterson, ``Roofline: an insightful visual performance model for multicore architectures,'' \emph{Communications of the ACM}, vol.~52, no.~4, pp. 65--76, April 2009. [Online]. Available: \url{https://doi.org/10.1145/1498765.1498785}
\BIBentrySTDinterwordspacing

\bibitem{lie2019introduction}
K.-A. Lie, \emph{{An introduction to reservoir simulation using MATLAB/GNU Octave: User guide for the MATLAB Reservoir Simulation Toolbox (MRST)}}.\hskip 1em plus 0.5em minus 0.4em\relax Cambridge University Press, 2019.

\bibitem{manea2015parallel}
A.~M. Manea, \emph{Parallel multigrid and multiscale flow solvers for high-performance-computing architectures}.\hskip 1em plus 0.5em minus 0.4em\relax Stanford University, 2015.

\bibitem{aziz}
K.~Aziz, ``Petroleum reservoir simulation,'' \emph{Applied Science Publishers}, vol. 476, 1979.

\bibitem{darcy1856fontaines}
\BIBentryALTinterwordspacing
H.~Darcy, \emph{Les fontaines publiques de la ville de Dijon: Exposition et application des principes {\`a} suivre et des formules {\`a} employer dans les questions de distribution d'eau : Ouvrage termin{\'e} par un appendice relatif aux fournitures d'eau de plusieurs villes, au filtrage des eaux et {\`a} la fabrication des tuyaux de fonte, de plomb, de t{\^o}le et de bitume}.\hskip 1em plus 0.5em minus 0.4em\relax Victor Dalmont, {\'e}diteur, 1856. [Online]. Available: \url{https://books.google.com/books?id=42EUAAAAQAAJ}
\BIBentrySTDinterwordspacing

\bibitem{groq22b}
D.~Abts, I.~Ahmed, A.~Bitar, M.~Boyd, J.~Kim, G.~Kimmell, and A.~Ling, ``{Challenges/Opportunities to Enable Dependable Scale-out System with Groq Deterministic Tensor-Streaming Processors},'' in \emph{{2022 52nd Annual IEEE/IFIP International Conference on Dependable Systems and Networks - Supplemental Volume (DSN-S)}}.\hskip 1em plus 0.5em minus 0.4em\relax IEEE, 2022, pp. 19--22.

\bibitem{DirkSC20}
K.~Rocki, D.~Van~Essendelft, I.~Sharapov, R.~Schreiber, M.~Morrison, V.~Kibardin, A.~Portnoy, J.~F. Dietiker, M.~Syamlal, and M.~James, ``{Fast Stencil-Code Computation on a Wafer-Scale Processor},'' in \emph{Proceedings of the International Conference for High Performance Computing, Networking, Storage and Analysis}, ser. SC '20.\hskip 1em plus 0.5em minus 0.4em\relax IEEE Press, 2020.

\bibitem{MauricioSC22}
M.~Jacquelin, M.~Araya-Polo, and J.~Meng, ``{Scalable Distributed High-Order Stencil Computations},'' in \emph{Proceedings of the International Conference on High Performance Computing, Networking, Storage and Analysis}, ser. SC '22.\hskip 1em plus 0.5em minus 0.4em\relax IEEE Press, 2022.

\bibitem{araya2023}
\BIBentryALTinterwordspacing
M.~Araya-Polo, M.~Jacquelin, and D.~Klahr, ``Massively distributed reverse time migration,'' vol. 2023, no.~1, pp. 1--5, 2023. [Online]. Available: \url{https://www.earthdoc.org/content/papers/10.3997/2214-4609.2023630010}
\BIBentrySTDinterwordspacing

\bibitem{ryuichi_scalah_2023}
R.~Sai, M.~Jacquelin, F.~P. Hamon, M.~Araya-Polo, and R.~R. Settgast, ``{Massively Distributed Finite-Volume Flux Computation},'' in \emph{{ScalAH23: 14th Workshop on Latest Advances in Scalable Algorithms for Large-Scale Heterogeneous Systems}}, 2023.

\bibitem{Samba21b}
M.~Emani, V.~Vishwanath, C.~Adams, M.~E. Papka, R.~Stevens, L.~Florescu, S.~Jairath, W.~Liu, T.~Nama, and A.~Sujeeth, ``{Accelerating Scientific Applications With SambaNova Reconfigurable Dataflow Architecture},'' \emph{Computing in Science \& Engineering}, vol.~23, no.~2, pp. 114--119, 2021.

\bibitem{louw2021using}
\BIBentryALTinterwordspacing
T.~Louw and S.~McIntosh-Smith, ``{Using the Graphcore IPU for traditional HPC applications},'' in \emph{{3rd Workshop on Accelerated Machine Learning (AccML), co-located with HIPEAC21}}, 2021. [Online]. Available: \url{https://workshops.inf.ed.ac.uk/accml}
\BIBentrySTDinterwordspacing

\bibitem{siam_tf_report}
\BIBentryALTinterwordspacing
B.~Hendrickson, A.~Aceves, E.~Alhajjar, M.~Bañuelos, D.~Brown, K.~Devine, Q.~Du, O.~Ghattas, R.~Giles, M.~Hall, T.~Islam, K.~Jordan, L.~Lin, A.~Pothen, P.~Raghavan, R.~Schreiber, C.~Thalhauser, and W.~Wilson, ``{SIAM Task Force Report: The Future of Computational Science},'' {Society for Industrial and Applied Mathematics}, Tech. Rep., Mar. 2024. [Online]. Available: \url{https://www.siam.org/Portals/0/Publications/Reports/SIAM%20Report%20on%20the%20Future%20of%20Computational%20Science.pdf?ver=2024-03-06-104353-590}
\BIBentrySTDinterwordspacing

\bibitem{Cerebras-Micro23}
S.~Lie, ``Cerebras architecture deep dive: First look inside the hardware/software co-design for deep learning,'' \emph{IEEE Micro}, vol.~43, no.~3, pp. 18--30, 2023.

\bibitem{yang_hierarchical_roofline_2020}
C.~Yang, T.~Kurth, and S.~Williams, ``{Hierarchical Roofline analysis for GPUs: Accelerating performance optimization for the NERSC-9 Perlmutter system},'' \emph{Concurrency and Computation: Practice and Experience}, vol.~32, no.~20, p. e5547, 2020, e5547 cpe.5547.

\bibitem{yang_roofline_ert_2020}
C.~Yang, ``{Empirical Roofline Toolkit},'' \textsc{url:}~\url{https://bitbucket.org/berkeleylab/cs-roofline-toolkit}, February 2021.

\end{thebibliography}

\end{document}